# The action as a differential *n*-form and the analytic deduction of the nuclear potentials.

Enrique Ordaz Romay[1]

*Facultad de Ciencias Físicas, Universidad Complutense de Madrid*

## Abstract

All the natural forces act through potential fields. Both, the electromagnetic vector potential and the gravitational potential of the general relativity are usually deduced starting from general analytic considerations.

However, the nuclear potentials of the weak and strong forces are calculated experimentally, leading to what is known as phenomenological potentials. That is to say, expressions requiring adjustments according to the experimental results.

This fact may change if we consider the action as a differential form. In this case it is possible to deduct some potentials for the nuclear forces departing from general analytic considerations.

---
[1] eorgazro@cofis.es

# Introduction

When a physical system is described by an action function as a differential form [1] [2], from the geometric standpoint the linear momentum corresponds to the components of such a differential form. In the 4-dimensional space-time the maximum rank (number of indexes) this differential form may have is four, due to the antisymmetry of its components. That is why, a decomposition in projections on a base leads to five terms, each one with a different rank and a very particular physical representation. We will dedicate the 1st part of this text to this analysis.

From the standpoint of the analytic physics each component of the previous representation defines a type of forces fields. This way, a rank-zero tensor, being a scalar, characterises a mechanical system without force fields, a rank-1 tensor, being a vector, characterises a field of electromagnetic type (the electromagnetic vector potential)[3]. A rank-2 tensor, that is to say, a matrix, characterises a field of the type of the weak forces (Fermi bilineal currents)[4], a rank-3 tensor would characterise a field of the strong force type [5] and, finally, a rank-4 tensor characterises a field of gravitational type (Riemman tensor) [3]. We will dedicate the second part of this paper to this analysis.

The quantum description of these fields is immediately obtained, departing from the generalisation of the complementarity principle [1]. We will take care of this in the 3rd part of this text.

# PART 1.
# Tensorial origin of the action function

When a system is described by an action vector function, its relationship with the traditional scalar action function is expressed by the equation [1]:

$$S^2 = S^i S_i$$

We need to know the form of the variation of the action in order to describe the space-time evolution of the physical system. From the previous expression we find (the differentiation operator $D$ corresponds to a derivative that, when applied on a tensor it results in another tensor):

$$2S \cdot dS = DS^j \cdot S_j + S_j g^{ji} \cdot DS_i$$

Since $g^{ij} DB_i = D(g^{ij} B_i) = DB^j$, becomes: $S \cdot dS = S_i \cdot DS^i$ or $S^i \cdot dS = S_. \cdot DS^i$ both of which directly lead to the expression:

$$(dS)^2 = DS^i \cdot DS_i \qquad (1)$$

This expression is completely similar to the one applied when the action has a tensor origin of arbitrary rank. In this case the expression gets the form:

$$(dS)^2 = DS^{ijk\ldots} \cdot DS_{ijk\ldots} \qquad (1')$$

This equation connects the variation of the scalar action with the variation of the tensor that originates the action to which we have called "action tensor".

## The action as function of the coordinates.

When we consider the scalar action function as a function of the coordinates, meaning $S = S(x^i)$, the form of the variation of the action, according to the rules of differentiation is:

$$dS = \frac{\partial S}{\partial x^i} dx^i = \partial_i S \cdot dx^i$$

To find the form of $DS^i$ we need to multiply the previous expression times its dual and to compare it with (1). By doing so we can see:

$$(dS)^2 = (\partial_i S \cdot dx^i) \cdot (\partial^i S \cdot dx_i) = \partial_i S \cdot \partial^i S \cdot (ds)^2 = (\partial_i S \cdot ds) \cdot (\partial^i S \cdot ds) \Rightarrow$$
$$DS^i = \partial^i S \cdot ds$$

In analytic mechanics the partial derivative of the action with respect to a coordinate is the negative of the momentum, that is to say: $\partial^i S = -P^i$. Therefore:

$$DS^i = -P^i ds \quad \Leftrightarrow \quad dS = -P^i dx_i \qquad (2)$$

Same way, we can calculate the second variation of the scalar action. Deriving the second equation of the expression (2) we have (the operator $D^i$ corresponds to the partial derivative which, applied on a *n* rank tensor, it produces a *n+1* rank tensor):

$$d^2 S = -(D^j P^i \cdot dx_j) \cdot dx_i - P^i d^2 x_i$$

Since, by perpendicularity $d^2 x_i = \partial^i (dx_i) dx_i = 0$ the former expression becomes:

$$d^2 S = -D^j P^i \cdot dx_i dx_j \quad \Leftrightarrow \quad D^2 S^i = -D^j P^i \cdot ds \cdot dx_j$$

that when compared with the expected expression $d^2 S = -P^{ij} \cdot dx_i dx_j$ it leads us to the relation:

$$P^{ij} = D^j P^i$$

That it matches the corresponding mechanical expression both in classical mechanics and in relativity [6]. Keeping in mind that *P* expresses the (square) norm of the linear momentum and its relationship with the momentum vector is $P^2 = P^i P_i$, a similar reasoning to the one leading to the expression (1) allows us to write:

$$P^{ij} P_{ij} = (dP)^2 \text{ Or in general } P^{ij\ldots} P_{ij\ldots} = (d^{n-1} P)^2$$

We could even find an expression for the differential operator $D$ that allows us to harmonize these relationships with the space-time evolution equation for the physical system according to the causality principle [2][7]. For it, using the notation of differential geometry [8], the form of the action as differential $n$-form will be:

$$d^n S = -P^{i_1 i_2 \dots i_n} dx_{i_2} \wedge dx_{i_2} \wedge \dots \wedge dx_{i_n}$$

$P$ being anti-symmetric to permutations of the indexes, that is to say: $P^{\dots i \dots j \dots} = -P^{\dots j \dots i \dots}$.

Applying this characteristic to the expression of $P^{ij}$ we are left with $D^j P^i = \frac{1}{2}\left(P^{i;j} - P^{j;i}\right)$ The superindex $i$ being the countervariant partial derivative according to the expression: $P^{i;j} = \partial^j P^i - g^{jl}\Gamma^i_{kl}P^k$ (being $\Gamma$ the Christoffel symbols)[3].

Using the notation of the differential geometry [8], according to which $C^{[ijk\dots]} = e^{ijk\dots}_{lmn\dots} C^{lmn\dots}$ being $e$ the anti-symmetric tensor that verifies the expression:

$$e^{ijk\dots}_{lmn\dots} = \begin{cases} +1 & \text{If } lmn\dots pqr\dots \text{ is even permutation of } ijk\dots \\ 0 & \text{If } lmn\dots pqr\dots \text{ or } ijk\dots \text{ have repeat index} \\ -1 & \text{If } lmn\dots pqr\dots \text{ is odd permutation of } ijk\dots \end{cases}$$

In summary: $P^{ij} = \frac{1}{2} P^{[i;j]}$ or in general: $P^{ijk\dots} = \frac{1}{n!} P^{[i;j;k;\dots]}$. (3)

These expressions are similar to those obtained from general considerations regarding the application of causality to the physical systems [2].

Since $d^n S$ is a differential $n$-form on a space with four dimensions, the highest value for $n$ is also 4. That is to say: the maximum number of indexes the linear momentum could have is four.

# Breakdown of the momentum
# in function of a given vector.

Let us suppose that we have a unitary vector $u$ ($u^i u_i = 1$). Given this vector we can always breakdown the $n$-tensor momentum according to its projections on the components of $u$, in the way:

$$P^{ijk...z} = \frac{1}{n!}\left(Pu^{[i}u^j u^k...u^{z]} + P^{[i}u^j u^k...u^{z]} + P^{[ij}u^k...u^{z]} + ... + C^{ijk...z}\right) \tag{4}$$

Being $C$ a tensor form of the perpendicular manifold to vector $u$.

When the momentum is a scalar, the expression (4) simplifies down to the identity $P = C$. Keeping in mind the symmetry of the product $u^i u^j = u^j u^i$, in the remaining cases we have:

$$n = 1: \quad P^i = Pu^i + C^i$$

$$n = 2: \quad P^{ij} = Pu^i u^j + \frac{1}{2}\left(P^i u^j - P^j u^i\right) + C^{ij}$$

$$n = 3: \quad P^{ijk} = Pu^i u^j u^k + \frac{1}{2}\left(P^i u^j - P^j u^i\right)u^k + \frac{1}{6}P^{[ij}u^{k]} + C^{ijk}$$

$$n = 4: \quad P^{ijkl} = Pu^i u^j u^k u^l + \frac{1}{2}\left(P^i u^j - P^j u^i\right)u^k u^l + \frac{1}{6}P^{[ij}u^{k]}u^l + \frac{1}{24}P^{[ijk}u^{l]} + C^{ijkl}$$

Below we will analyse some physical systems using this equations. Each system has one of the fundamental forces. We will observe that all fundamental forces can be described by an action expressed as differential $n$-form.

# PART 2

# Analysis of physical systems.

**a) homogeneous and isotropic systems.**

A homogeneous and isotropic physical system is characterised totally by a scalar physical magnitude [7]. Therefore the action will be a linear function in the interval *s*. That is to say: $S = -P \cdot s$.

The scalar magnitude that characterises the homogeneous and isotropic system is usually called mass, for this reason $P \propto m$. The proportionality constant between *P* and *m*, with units of speed, should then be a constant value in any inertial frame of reference. This value can only be the speed of light *c*. Therefore, our physical system is characterised by the following expression of the action:

$$S = -mc \cdot s \quad \Leftrightarrow \quad P = mc \tag{5}$$

**b) System composed by a single free particle.**

Although, in a strict sense a particle having mass cannot be free, since the non homogeneous distribution of mass in the system, due to the presence of the particle, leads by itself to an alteration in the metric of the system, we will consider, for the moment, that such alteration is negligible in view of the magnitudes that we will deal with.

Since we have characterised the system by the presence of a "free" particle with idle mass *m*, this particle will characterise a direction of the space as preferable. This direction will correspond to the variation in the position of the particle when it moves from a point *X* to the point *X + dX*. Since the system is free, such variation can only be referred to the interval.

By representing the variation in the coordinate $x^i$ by $dx^i$, we define the unitary vector $u^i = \dfrac{dx^i}{ds}$ as the 4dimensional speed. It is easy to see that its form is:

$$u^i = \frac{1}{\sqrt{1-\dfrac{v^2}{c^2}}}\left(1 \quad \frac{v_x}{c} \quad \frac{v_y}{c} \quad \frac{v_z}{c}\right)$$

Being ($v_x$, $v_y$, $v_z$) the traditional three-dimensional speed vector. By applying the equation (2) we get $P^i = Pu^i$. As by hypothesis we have considered that the presence of the particle does not break the homogeneity and isotropy of the system (its effects are negligible), knowing that $P = mc$ we obtain the expression:

$$dS = -P^i dx_i \quad \Leftrightarrow \quad P^i = mcu^i \tag{6}$$

Or also: $dS_i = -P dx_i \quad \Leftrightarrow \quad P = mc$

This tells us that, when we have a free particle, the 4-speed of the particle characterises the system. That is to say, the unit vector $u$ shows us a direction regarding which we can decompose or break our system down.

When the particle is only characterised by its scalar magnitude (its mass), the linear momentum, as a description of the system can only be a function of the 4-speed and of the mass. Under these conditions the system is called mechanic and all its properties, such as the conservation principles or its movement equations are deduced from the equation:

$$P^i P_i = m^2 c^2$$

When we substitute $P_i$ by its traditional expression $-\partial^i S$ we obtain the *equations of Hamilton-Jacobi*.

## c) System composed by a particle in a vector field. The electromagnetic field.

A vector field is characterised by a vector potential having the form $C^i$. Following the breakdown (4), for $n = 1$ the momentum can be decomposed, according to the 4-speed vector in the form:

$$P^i = Pu^i + C^i \tag{7}$$

On the other hand, for $n = 2$ we have $d^2S = DS_{ij}DS^{ij}$ being:

$$DS^{ij} = -\frac{1}{2}C^{[i;j]}(ds)^2 \quad \Rightarrow \quad (dS)^2 = \frac{1}{4}C^{[i;j]}C_{[i;j]}dx^i \wedge dx^j \wedge dx_i \wedge dx_j \tag{8}$$

Since in the 4-space the maximum number of different indexes is 4, the form:

$$(dx^i \wedge dx^j) \wedge (dx_i \wedge dx_j) = g^{ik}g^{jl}dx_i \wedge dx_j \wedge dx_k \wedge dx_l = \sqrt{-g}\,e^{0123}_{ijkl}dx_0 dx_1 dx_2 dx_3$$

Is equal to $\sqrt{-g}\,d\Omega$. This represents the Lorentz invariant 4-volume element.

Keeping this in mind, we may rewrite the expression (8) in the form:

$$dS = \frac{1}{4}C^{[i;j]}C_{[i;j]}\sqrt{-g}\,d\Omega \tag{9}$$

From the expressions (5) and (7) we may say that, in presence of a vector field, the action adopts the form:

$$dS = -Pds - C^i dx_i - \frac{1}{4}C^{[i;j]}C_{[j;i]}\sqrt{-g}\,d\Omega \tag{10}$$

When the system is formed by a vector field, just as in the case of an electromagnetic field, observing that the electromagnetic vector potential $A^i$ must be proportional to the vector $C^i$. Being the constant of proportionality the same vector magnitude that characterises the particle. We have $C^i \propto eA^i$, being $e$ what is known as the electric charge of the particle.

Since $C$ has units of Mass $\times$ Speed, while $eA$ has units of Energy, that is to say, Mass $\times$ Speed$^2$, the proportionality constant should have units of Speed$^{-1}$. Given the independence of the reference system, the only constant that we can use is $1/c$. Therefore:

$$C^i = \frac{e}{c} A^i$$

On the other hand being so that $C^{[i;j]} = C^{i;j} - C^{j;i} = \frac{e}{c}\left(A^{i;j} - A^{j;i}\right)$. This allows us to define the tensor of rank 2: $F^{ij} = A^{i;j} - A^{j;i}$ which is called: tensor of the electromagnetic field.

Applying these results to the relationship (10) we are left with:

$$dS = -mc \cdot ds - \frac{e}{c} A^i dx_i - \frac{1}{4} F^{ij} F_{ij} \sqrt{-g}\, d\Omega$$

This expression exactly matches the *equation of the electrodynamics* [3].

## d) System for a rank 4 field tensor. Gravitational field.

For $n = 4$, we will assume that, the only part of the linear momentum that is different from zero is (in $n = 4$ there is only a collection of 4-indexes linearly independent): $P^{ijkl} = P u^i u^j u^k u^l + C^{ijkl}$. To guarantee the momentum components asymmetry we can contract the indexes to the form:

$$P = P^{ik} g_{ik} - C^{ijkl} g_{ik} g_{jl}$$

This equation may be rewritten as:

$$C_{ik} = P_{ik} - Pg_{ik} \tag{11}$$

From this, it is easy to obtain the equations of the gravitational field by rewriting it in a more familiar way. In General Relativity, the density of the 4-tensor field is called *tensor of Riemann*. It is denoted by: $R_{ijkl}$ and its contraction, expressed as $R_{ik}$ is the *tensor of Ricci*, Due to this, the tensors $R_{ik}$ and the 4-density of $C_{ik}$ must be proportional to each other.

On the other hand, the energy-momentum tensor $T_{ik}$ commonly used in the classic theory of fields [3] has the form $T_{ik} = -\dfrac{\partial \Lambda_k}{\partial x^i}$, being: $\Lambda_k = \Lambda \cdot u_k$ ($\Lambda$ is the 4-dimensional density of the scalar action, or the same thing as the three-dimensional density of the lagrangian). According to this, the tensor $T_{ik}$ is the density of the tensor $P_{ik}$.

Taking into account the former two reasonings, the equation (11) becomes:

$$R_{ik} \propto T_{ik} - Tg_{ik}$$

The proportionality constant will be such that, for small values of $R$ this equation matches the Newton's expression of the gravitation. Making this approximation we find the constant $\dfrac{8\pi k}{c^4}$, being $k$ Newton's gravitation constant. Substituting we obtain:

$$R_{ik} = \dfrac{8\pi k}{c^4}\left(T_{ik} - Tg_{ik}\right)$$

Which are in fact the *equations of the gravitational field*, also called *Einstein's equations*.

# Nuclear forces: Weak and strong fields.

Although, the analysis we have made up to this point has not taken into account the mathematical realisation. It is certain that the description of the weak and strong forces only makes full sense inside a quantum framework.

However, the field equations for the linear momentum expressed as differential *n*-form lead us to tensor components of 2nd and 3rd order whose expressions may match with those of the strong and weak force field types.

The application of the principle of correspondence into these equations [1], as well as a mathematical-interpretative quantum-relativistic analysis will lead our field equations to the correct expressions corresponding to the nuclear forces.

## e) System composed by a matrix field. Weak field.

Based on expression (8) we can observe that, as $C^{ij}C_{ij}$ is an invariant, we can find a vector $J$ such that $C^{ij}C_{ij}=J^a J_a$. This vector $J$ is traditionally called 4-vectorial weak current, and due to (11) $dS \propto \int J^\alpha J_\alpha$.

The most peculiar characteristic of the weak force is the non-conservation of the symmetry of parity [9].

The system formed by the matrix field is characterized by: $C^{ij}$. In presence of a particle the expression of the action should have the form of the equation (8), plus an additional term due to the matrix component. We will concentrate on this last term.

As: $P^{ij}P_{ij} = const$ then: $C^{ij}C_{ij} = \frac{1}{4}\left[P^{i;j} - P^{j;i}\right]\cdot\left[P_{i;j} - P_{j;i}\right]$.

Developing we obtain:

$$C^{ij}C_{ij} = \frac{1}{4}\left[2\left(P^{i;j}P_{i;j}\right) - \left(P^{j;i}P_{i;j} + P^{i;j}P_{j;i}\right)\right] = \frac{1}{4}\left[V - A\right] \qquad (12)$$

Being $V = 2(P^{i;j} P_{i;j})$ the squared module of an object that transforms into a tensorial one, while $A = P^{j;i} P_{i;j} + P^{i;j} P_{j;i}$ is a pseudo scalar that corresponds to the squared module of an object that transforms as a pseudo tensor.

Now then, up to this point we have made reference to a property of the matrix field. In order to calculate its potential, starting with (4) we find:

$$P^j = Pu^j + \frac{1}{6} g_{ik} P^{[ij} u^{k]} \Rightarrow \frac{1}{6} g_{ik} P^{[ij} u^{k]} u_j = 0 \Rightarrow$$

$$\Rightarrow \frac{1}{6}\left[2 P^{[ij]} u^i u^j + P^{[ki]} g_{ik}\right] = 0$$

This last one is the field equation we have been looking after for the weak force, subject to the restriction (11).

## f) System formed by a rank 3 tensorial field. Strong field.

Assuming $n = 3$ and the rest of the components of the momentum to be zero, the expression (4) gets the form:

$$P^{ijkl} = Pu^i u^j u^k u^l + \frac{1}{24} P^{[ijk} u^{l]} \Rightarrow P^{ik} = Pu^i u^k + \frac{1}{24}\left(g_{jl} P^{[ijk} u^{l]}\right)$$

Contracting this expression we get: $g_{jl} \frac{1}{24}\left(P^{[ijk} u^{l]}\right) g_{ik} = 0$.

Operating we obtain:

$$g_{jl} \frac{1}{24}\left(P^{[ijk} u^{l]}\right) g_{ik} = 0 \Rightarrow u_i u_j \frac{1}{24}\left(P^{[ijk} u^{l]}\right) u_k u_l = 0 \Rightarrow$$

$$\Rightarrow \frac{1}{24}\left[u_i u_j P^{ijk} u_k - u_i u_j P^{ijl} u_l + u_i u_k P^{ikl} u_l - u_i u_j P^{jik} u_k + u_i u_j P^{jil} u_l - u_j u_k P^{jkl} u_l ...\right] = 0$$

$$\Rightarrow \frac{1}{24}\left[u_i u_j P^{[ijk]} u_k + u_i u_j P^{[ijl]} u_l + u_i u_k P^{[ikl]} u_l + u_j u_k P^{[jkl]} u_l\right] = 0$$

This one is the field equation looked after for the strong force.

# PART 3.

# Quantum-relativistic extension

We could not complete a description of the nuclear forces, or a complete description of the electromagnetic and gravitational fields if we wouldn't make a quantum-relativistic extension for our field equations.

In quantum-relativistic physics the described objects are expressed by means of spinorial wave functions. Instead of Dirac's spinorial notation, we will here follow the spinor notation corresponding to the co-variant transformations [1]. In this notation the wave function mathematical object verifies the relationship: $\Psi^i \Psi_i = 1$.

Departing from the expression of the metric $dx^i dx_i = (ds)^2$ and reminding the expression of the 4-speed as: $u^i = \frac{dx^i}{ds}$, we obtain $u^i u_i = 1$. The analogy between this expression and the relationship for the wave function leads us to the transformation $u^i \rightarrow \Psi^i$.

On the other hand, the generalization of the principle of complementarity [1] allows us to find the transformation for the linear momentum: $P^{ij} \rightarrow -i\hbar D^i \Psi^j$.

Applying these two transformations to the field equations we obtain the quantum-relativistic extension. This way, applying it to (6) we obtain:

$$-i\hbar D \Psi^j = mc \Psi^j$$

This is *Dirac's equation* [10] in covariant notation.

Likewise, the expression for the strong force: $u_i u_j P^{[ijk]} u_k = 0$ transforms in: $-\hbar^2 \Psi_i^{(1)} \Psi_j^{(2)} D^{[i} D^j \Phi^{k]} \Psi_k^{(3)} = 0$.

Finally, for the gravitational field case, we should find an equivalent relationship to the one of the principle of complementarity, that transforms for us the metric into wave functions. On that purpose, being that:

$$\Psi_i g^{ij} \Psi_j = 1$$

Starting with (11) we arrive to:

$$R_{ik} \propto -i\hbar \left( D_i \Psi_k - \Psi_k D_l \Psi^l \Psi_i \right)$$

This equation connects the Riemman´s tensor (the space-time deformation) and the wave function. That is to say, the wave function deforms the space-time. This result is evident when we consider that, according to general relativity, the mass deforms the space and that, according to quantum mechanics, every particle is described by a wave function. Consequently, the very wave function deforms the space-time. The last equation says exactly that same thing.

## Acknowledgement:

Thanks to Jorge Millán Ayala for taking interest in this work and for the help he provided with the English language by revising the last version of this paper.

## Bibliography.